\documentclass[twocolumn,prb,floatfix]{revtex4}
 
\usepackage{amsmath}
\usepackage{amssymb}
\usepackage{graphicx}
\usepackage{subfigure}
\usepackage{float}
\usepackage{graphics}
\usepackage{epsfig}

\begin{document}

\title{Optical and magneto-optical far-infrared properties of bilayer graphene}
\author{D. S. L. Abergel and Vladimir I. Fal'ko}
\affiliation{Physics Department, Lancaster University, Lancaster, LA1 4YB, UK}

\begin{abstract}
We analyze the spectroscopic features of bilayer graphene determined by the
formation of pairs of low-energy and split bands in this material. We show that
the inter-Landau-level absorption spectrum in bilayer graphene at high magnetic
field is much denser in the far-infrared range than that in monolayer material, and that
the polarization dependence of its lowest-energy peak can be used to test the
form of the bilayer ground state in the quantum Hall-effect regime.
\end{abstract}

\maketitle

Monolayer and bilayer graphenes\cite{Novoselov,Zhang,QHEbi}
are gapless two-dimensional (2D) semiconductors \cite{AndoReview,McCann}.
When used in transistor-type devices, the density and type ($n$ or $p$) of
carriers in them can be controlled using the underlying gate \cite{Novoselov},
which has been exploited in the recent studies of the quantum Hall effect
(QHE) in such structures. The sequencing of QHE plateaus in graphene
\cite{Novoselov,Zhang,QHEbi} revealed the peculiar properties of the charge
carriers in this material: Dirac-type chiral electrons with Berry phase $\pi$
in monolayers \cite{QHE,AndoReview}, and the Berry phase $2\pi$ chiral
quasiparticles \cite{McCann} with doubled degeneracy of the zero-energy Landau
level (LL) in bilayers.

The electromagnetic (EM) field absorption in graphene at zero magnetic field
has already been studied \cite{GusSharMW,Nilsson,Varlamov,Czerti}. While the
dc conductivity of monolayer graphene increases linearly with the carrier
density \cite{Novoselov,NomuraMacDonald}, the real part of its high-frequency
conductivity \cite{GusSharMW,Varlamov} is independent of the electron density
in a wide spectral range above the threshold $\hbar \omega
>2|\epsilon _{\mathrm{F}}|$, which determines a featureless absorption
coefficient $g_{1}=\pi e^{2}/\hbar c$. In contrast, the bilayer absorption
coefficient, 
\begin{equation}
	g_{2}=(2\pi e^{2}/\hbar c)\,f_{2}(\omega ),  \label{Eq-1}
\end{equation}
reflects the presence and dispersion of two pairs of bands in this material 
\cite{McCann,Nilsson,Eli,Others}. In this paper, we analyse far-infrared
(FIR) magneto-optical properties due to the inter-LL \cite{mcc56,CR,McCann}
transitions in bilayer graphene and compare them to those in the monolayer
material. The low-energy part of the absorption spectrum in a bilayer
subjected to a strong perpendicular magnetic field is much denser than that of
a monolayer, reflecting the parabolic dispersion of its low-energy bands, with
the lowest FIR absorption peak formed by transitions involving one of two
degenerate LLs at zero energy. Although the high-energy part of the bilayer
absorption spectrum is not sensitive to the polarisation of FIR irradiation,
below we show that its lowest peak may be strongly polarised reflecting the
occupancy of the degerate LLs in a bilayer determined by the form of its
ground state in the QHE regime.

The electronic Fermi line in graphene surrounds the corners \cite{kpoints} 
$\mathbf{K}_{\pm }$ of the hexagonal Brillouin zone \cite{AndoReview} (where
we set $\epsilon =0$). In monolayer graphene, quasiparticles near the
centres of valleys $\mathbf{K}_{\pm }$ can be described by 4-component Bloch
functions $\psi ={[}\phi _{\mathbf{K}_{+}A}$, $\phi _{\mathbf{K}_{+}B}$, 
$\phi _{\mathbf{K}_{-}B}$, $\phi _{\mathbf{K}_{-}A}]$ and the Hamiltonian 
$\hat{H}_{1}= \upsilon \Pi_z \otimes \boldsymbol{\sigma} \cdot
\mathbf{p}$, where
$\boldsymbol{\sigma} =(\sigma _{x},\sigma _{y})$ are Pauli matrices acting in
the space of electronic amplitudes on the two crystalline sublattices ($A$ and
$B$) and $\Pi_z$ is the diagonal Pauli matrix in the valley space. Momentum
$\mathbf{p=}-i\hbar \mathbf{\nabla }-\frac{e}{c}\mathbf{A}$ is calculated with
respect to the center of the corresponding valley and $\mathbf{\nabla\times
A}=B\mathbf{l}_z$.

\textit{Bilayer graphene} is composed of two coupled monolayers (with
sublattices $A,B $ and $\tilde{A},\tilde{B}$ in the bottom and top layers
respectively) arranged according to Bernal stacking \cite{AndoReview}: sites 
$B$ of the honeycomb lattice in the bottom layer lie below $\tilde{A}$ of
the top layer. It also has a hexagonal Brillouin zone with two inequivalent
valleys, but carries twice the number of electronic dispersion branches.
The latter can be found using the nearest-neighbour hopping Hamiltonian
\cite{McCann} acting in the space of sublattice states $[\phi
_{\mathbf{K}_{+}A}$, 
$\phi _{\mathbf{K}_{+}\tilde{B}}$, $\phi _{\mathbf{K}_{+}\tilde{A}}$, 
$\phi_{\mathbf{K}_{+}B}$; $\phi _{\mathbf{K}_{-}\tilde{B}}$, 
$\phi_{\mathbf{K}_{-}A}$, $\phi _{\mathbf{K}_{-}B}$, 
$\phi_{\mathbf{K}_{-}\tilde{A}}]$, 
\begin{equation}
	\hat{\mathcal{H}}_{2} =
	\begin{pmatrix}
		\upsilon_{3}\Pi_z\otimes\boldsymbol{\sigma}^{\mathrm{t}} \cdot
			\mathbf{p} & 
		\upsilon \Pi_z \otimes \boldsymbol{\sigma}\cdot \mathbf{p} \\ 
		\upsilon \Pi_z \otimes \boldsymbol{\sigma} \cdot \mathbf{p} & 
		\gamma_{1} \Pi_0 \otimes \sigma_{x}
	\end{pmatrix}.
	\label{bilayer-full}
\end{equation}
Here, $\upsilon $ is determined by the $AB$ ($\tilde{A}\tilde{B}$)
intra-layer hopping, $\gamma _{1}$ is the strongest inter-layer $\tilde{A}B$
hopping element, and $\upsilon _{3}\ll \upsilon $ is due to a weak direct 
$A\tilde{B}$ hop, and `t' stands for transposition.  Equation
(\ref{bilayer-full}) determines two types of branches in the electronic
spectrum of graphene \cite{McCann,Others} observed in the recent ARPES studies
\cite{Eli}: Split-bands, 
\begin{equation}
	\varepsilon _{s}^{\pm }=\pm (\gamma _{1}/2{)}
	[\sqrt{1+4\upsilon^{2}p^{2}/\gamma _{1}^{2}}+1],  \label{Es}
\end{equation}
formed by symmetric and antisymmetric states based upon $\tilde{A}B$
sublattices; and two gapless branches, $\varepsilon _{c}^{\pm }$. For 
$\epsilon \gg \epsilon _{\mathrm{L}}\equiv \frac{1}{2}\gamma _{1}(\upsilon
_{3}/\upsilon )^{2}$, the dispersion in gapless branches can be approximated
by 
\begin{equation}
	\varepsilon _{c}^{\pm }=\pm (\gamma _{1}/2{)}
	[\sqrt{1+4\upsilon^{2}p^{2}/\gamma _{1}^{2}}-1].  \label{Ec}
\end{equation}
For $|\epsilon |<\frac{1}{4}\gamma _{1}$, gapless bands are formed by states
from sublattices $A$ and $\tilde{B}$ described using the 4-component Bloch
functions $\chi ={[}\phi _{\mathbf{K}_{+}A}$, $\phi_{\mathbf{K}_{+}\tilde{B}}$,
$\phi _{\mathbf{K}_{-}\tilde{B}}$, $\phi_{\mathbf{K}_{-}A}]$ and the reduced
low-energy Hamiltonian \cite{McCann,McCann-new}, 
\begin{gather}
	\hat{H}_{2}=-\frac{1}{2m_{2}} (\boldsymbol{\sigma} \cdot \mathbf{p}) \sigma_{x} 
		(\boldsymbol{\sigma} \cdot \mathbf{p}) + \delta \hat{h}_{w} 
		+ \frac{\beta p^{2}}{m_{2}} ; \notag \\
	\delta \hat{h}_{w} = \upsilon_3 \Pi_z \otimes 
		\boldsymbol{\sigma}^{\mathrm{t}} \cdot \mathbf{p}; \quad
		m_{2}=\gamma_{1}/2\upsilon ^{2},\;|\beta |\ll 1. \label{eq:Ham2}
\end{gather}
Here $m_2 \approx 0.05 m_e$ and the last term in $\hat{H}_{2}$ takes into
account the weak $AA$ and $BB$ intra-layer hopping \cite{beta}.

In a 2D electron gas with conductivity $\sigma(\omega)$ much less than
$c/2\pi$, absorption of an EM field $\mathbf{E}_{\omega }=\boldsymbol{\ell}
Ee^{-i\omega t}$ with polarisation $\boldsymbol{\ell}$ ($\boldsymbol{\ell}
_{\oplus }=[\mathbf{l}_{x}-i\mathbf{l}_{y}]/\sqrt{2}$ for right-
and $\boldsymbol{\ell }_{\ominus }=[\mathbf{l}_{x} +
i\mathbf{l}_{y}]/\sqrt{2}$ for left-hand circularly polarised light,respectively arriving
along the direction antiparallel to a magnetic field) can be characterised by
the absorption coefficient $g\equiv E_{i}E_{j}^{\ast }\sigma _{ij}(\omega
)/S$: the ratio between Joule heating and the energy flux $\mathbf{S}=
c\mathbf{E\times H}/4\pi =-S\mathbf{l}_{z}$ transported by the EM field. Using
the Keldysh technique, we express 
\begin{equation*}
	g=\frac{8e^{2}}{c\omega }\Re \int \frac{Fd\epsilon }{N}\widehat{\mathrm{Tr}}
	\left\{ \hat{\upsilon}_{i}\ell _{i}\hat{G}^{R}(\epsilon )\hat{\upsilon}
	_{j}\ell _{j} \hat{G}^{A}(\epsilon +\omega )\right\} ,
\end{equation*}
where $\mathbf{\hat{v}}=\partial _{\mathbf{p}}\hat{H}$ is the velocity operator,
$\widehat{\mathrm{Tr}}$ includes the summation both over the sublattice indices
``tr" and over single-particle orbital states, $N$ is the normalisation area of
the sample and $F=n_{\mathrm{F}}(\epsilon )-n_{\mathrm{F}}(\epsilon +\omega )$
takes into account the occupancy of the initial and final states. Here we have
included spin and valley degeneracy.

\begin{figure}[tb]
\centering
\includegraphics[clip,scale=0.48]{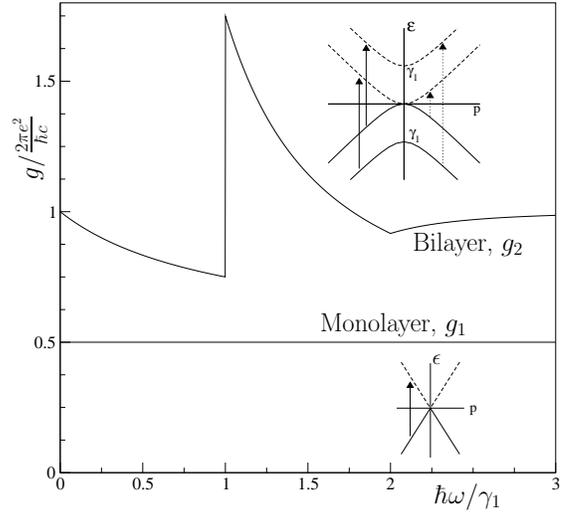}
\caption{Absorption coefficient of bilayer and monolayer graphene in the
optical range of frequencies. Insets illustrate the quasiparticle dispersion
branches in the vicinity of $\protect\epsilon _{F}$ and possible optical
transitions.}
\label{fig:AC-zero-field}
\end{figure}

For a 2D gas in a zero magnetic field, the electron states are weakly
scattered plane waves. Using the plane wave basis and the matrix form of the
high energy bilayer Hamiltonian $\hat{\mathcal{H}}_{2}$, we express the
retarded/advanced Greens functions of electrons in the bilayer as $\hat{G}
^{R/A}\left( \mathbf{p},\epsilon \right) =\left[ \epsilon \pm \tfrac{1}{2}
i\hbar \tau ^{-1}-\hat{H}_{2}(\mathbf{p})\right] ^{-1}$ and $\widehat{
\mathrm{Tr}}=\int d^{2}\mathbf{p}\frac{N}{\left( 2\pi \hbar \right) ^{2}}
\mathrm{tr}$, neglecting the renormalisation of the current operator
by vertex corrections at $\omega \tau \gg 1$ and the momentum transfer from
light (since $\upsilon /c\sim 3\times 10^{-3}$). This reproduces the constant
absorption coefficient $g_{1}^{\parallel}=\pi e^{2}/ \hbar c$ so that
$f_{1}=\frac{1}{2}$ \cite{GusSharMW,Varlamov} in monolayer graphene and
yields the following expression for the absorption coefficient of bilayer
graphene for light polarised in the plane of its sheet \cite{transv}: 
\begin{gather}
	g_{2}^{\parallel }=\frac{2\pi e^{2}}{\hbar c}f_{2}(\Omega ),\qquad \Omega
		\equiv \frac{\hbar \omega }{\gamma _{1}}>\frac{2|\epsilon _{\mathrm{F}}|}{
		\gamma _{1}}, \\ 
	f_{2}=\frac{\Omega +2}{2(\Omega +1)}+\frac{\theta (\Omega -1)}{\Omega ^{2}}+
	\frac{(\Omega -2)\theta (\Omega -2)}{2(\Omega -1)}.  \label{g2} 
\end{gather}
Here $\theta (x<0)=0$ and $\theta (x>0)=1$. This result agrees with
the calculation by J. Nilsson \textit{et al} \cite{Nilsson} taken in the
clean limit and at $T=0$. The frequency dependence \cite{frequency} of the
bilayer optical absorption is illustrated in Fig.\ref{fig:AC-zero-field},
where an additional structure in the vicinity of $\hbar \omega =\gamma _{1}$
($\gamma _{1}\approx 0.4$eV (Refs. \cite{AndoReview} and \cite{Eli}) is due to the
electron-hole excitation between the low-energy band $\varepsilon _{c}^{\pm }
$ and the split band $\varepsilon _{s}^{\pm }$. For the higher photon
energies, $\hbar \omega \gtrsim 2\gamma _{1}$, the frequency dependence
almost saturates at $f\approx 1$. Over the entire spectral interval shown in
Fig.\ref{fig:AC-zero-field}, the absorption coefficient for the left- and
right-handed light are the same, so that Eq. (\ref{g2}) is applicable \cite
{transv} to light linearly polarised in the graphene plane.

\textit{High-field FIR magneto-optics of graphene. }In a magnetic field, the
continuous zero-field spectrum, [Eq. (\ref{g2})] splits into lines determined
by transitions between Landau levels. The LLs can be studied after rewriting 
$\hat{H}_{1}$ and $\hat{H}_{2}$ in terms of descending, ${\pi }={p}
_{x}+i{p}_{y}$ and raising, ${\pi }^{\dagger }={p}_{x}-i{p}_{y}$ operators
in the basis of Landau functions $\varphi _{n\geq 0}$, leading to 
\begin{equation*}
	\boldsymbol{\sigma \cdot p} = 
	\left( \begin{array}{cc} 0 & {\pi }^{\dag } \\ {\pi } & 0 \end{array} \right).
\end{equation*}
The resulting monolayer spectrum \cite{mcc56} contains 4-fold
degenerate (2$\times $ spin and 2$\times $ valley index) states: one at $
\epsilon _{0}=0$ with $\psi _{0K_{+}}=[\varphi _{0},0,0,0]$ and $\psi
_{0K_{-}}=[0,0,\varphi _{0},0]$, and pairs of levels $\epsilon _{n\pm }=\pm
\hbar \upsilon \lambda _{B}^{-1}\sqrt{2n}$ ($\lambda _{B}=\sqrt{\hbar c/eB}$
is the magnetic length) with $\psi _{n\pm ,K_{+}}=\frac{1}{\sqrt{2}}[\varphi
_{n},\pm i\varphi _{n-1},0,0]$ and $\psi _{n\pm ,K_{-}}=\frac{1}{\sqrt{2}}
[0,0,\varphi _{n},\mp i\varphi _{n-1}]$. We neglect the electron spin
splitting and use the index $\alpha =\pm $ in $\epsilon _{n\alpha }$ and $
\psi _{n\alpha ,K}$ for the conduction ($+$) and valence ($-$) band LLs.

The bilayer spectrum features \cite{McCann} a group of eight states with 
$\varepsilon _{0}\approx \varepsilon _{1}\approx 0$  with wave functions 
$\chi_{0,K_{+}}=[\varphi _{0},0,0,0]$, $\chi _{0,K_{-}}=[0,0,\varphi _{0},0]$ and
$\chi_{1,K_{+}}=[\varphi _{1},0,0,0]$, $\chi _{1,K_{-}}=[0,0,\varphi _{1},0]$ 
(4$\times $2, due to spin degeneracy), and a ladder of almost equidistant
4-fold degenerate levels $\varepsilon _{n\pm }=\pm \hbar \omega_{c}
\sqrt{n(n-1)}$ (2$\times $ spin and 2$\times $ the valley index) with $\hbar
\omega _{c}=\hbar ^{2}/m_{2}\lambda _{B}^{2}$ and wave functions $\chi
_{n\pm ,K_{+}}=\frac{1}{\sqrt{2}}[\varphi _{n},\pm \varphi _{n-2},0,0]$, 
$\chi_{n\pm ,K_{-}}=\frac{1}{\sqrt{2}}[0,0,\varphi _{n},\pm \varphi _{n-2}]$.
The latter result can be found from the analysis of the first term in the
low-energy-band Hamiltonian $\hat{H}_{2}$, which dominates for $|\epsilon | <
\frac{1}{4}\gamma _{1}$ and high magnetic fields such that $\lambda
_{B}^{-1}>\gamma _{1}\upsilon _{3}/\upsilon ^{2}$. Numerical diagonalisation
of the full $\hat{H}_{2}$ and $\hat{\mathcal{H}_{2}}$ shows \cite{McCann}
that this degeneracy is not lifted by the warping term $\delta \hat{H}_{w}$,
and also the above-described grouping of LLs in bilayer graphene was
confirmed in the recent QHE measurements \cite{QHEbi}. It has been noticed 
\cite{McCann-new} that degenerate levels with $n=0$ ($\chi _{0,K_{+}},\chi
_{0,K_{-}}$) and $n=1$ ($\chi _{1,K_{+}},\chi _{1,K_{-}}$) can be weakly split
by the second-neighbour hopping term $\beta p^{2}/m_{2}$, to $\varepsilon
_{0}=\frac{\beta }{2}\hbar \omega _{c}$ and $\varepsilon _{1}=
\frac{3\beta}{2}\hbar \omega _{c}$, where $\beta \ll 1$ \cite{beta}. However,
it is more likely that, at a high magnetic field, electrons in an ideally
clean bilayer with filling factor $-4<\nu <4$ would form a correlated ground
state in which the occupancy of $n=0$ and $n=1$ LLs would be determined by the
electron-electron interaction. The correlated ground state may be particularly
interesting in a bilayer with $\nu =0$. One can envisage 2D electrons forming
a ferromagnetic QHE state in which both $n=0$ and $n=1$ states are half-filled
(like in a convensional QHE liquid at the filling factors corresponding to a
half-filled spin-degenerate LL).
Alternatively, there can be an antiferromagnetic state with one fully occupied
and one empty LL.  Below we show that these two ground states can be
distinguished experimentally on the basis of magneto-absorption spectra
measured in circularly polarised FIR light.

\begin{figure}[tb]
\includegraphics{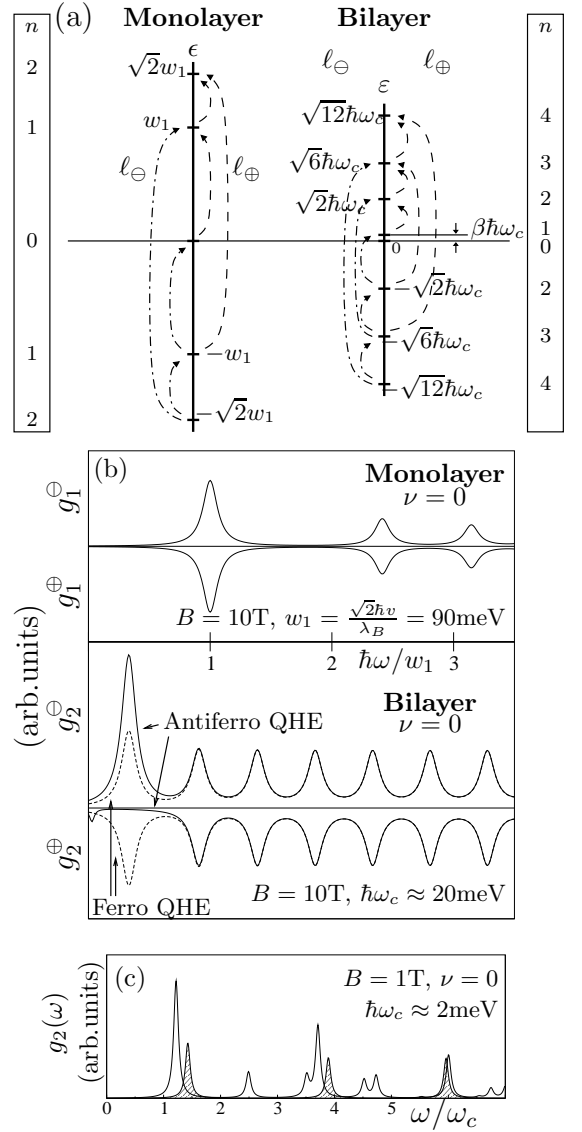}
\caption{(a) Allowed inter-LL transitions without trigonal warping effects.
Dashed and dash-dot lines indicate transitions in $\ell _{\oplus }$ and $
\ell _{\ominus }$ polarisations respectively. (b) Monolayer (top) and
bilayer (bottom) FIR absorption spectra in $\ell _{\oplus }$ and $\ell
_{\ominus }$ polarisations for $B=10T$ and filling factor $\protect\nu =0$.
Dashed and solid lines describe absorption by ferro- and antiferromagnetic
states of the $\protect\nu =0$ bilayer. (c) Weak field, $B=1$T
magnetoabsorption in bilayer graphene with $\protect\nu =0$ calculated with (
$\protect\upsilon _{3}=0.2\protect\upsilon $) and without ($\protect\upsilon 
_{3}=0$, shaded) warping term in the
Hamiltonian.\label{fig:transitions} }
\end{figure}

Figure \ref{fig:transitions}(a) illustrates the selection rules for the
single-particle inter-LL transitions in bilayer and monolayer graphene. In
both cases, photons with in-plane \cite{transv} polarisation
$\boldsymbol{\ell}_{\ominus }$ are absorbed via transitions where the LL index
changes from $n$ to $n-1$ ($n\geq 1$), whereas absorption of
$\boldsymbol{\ell}_{\oplus}$ photons happens via transitions from the LL $n$
to $n+1$ ($n\geq 0$).  Assuming the same broadening $\hbar \tau ^{-1}$ of all LLs
\cite{broadenning}, we arrive at the following magneto-absorption spectra for
monolayer \cite{CR} and bilayer graphene (for $\epsilon _{\mathrm{L}}<\hbar
\omega <\frac{1}{4}\gamma _{1}$): 
\begin{gather}
	g_{J}^{\oplus /\ominus }(B,\omega )=\frac{2\pi e^{2}}{\hbar c}f_{J}^{\oplus
		/\ominus }(B,\omega ),  \label{g-magn} \\
	f_{1}^{\ominus }=\sum_{\substack{ n\geq 1  \\ \alpha \alpha ^{\prime }}}
		\frac{\frac{w_{1}\tau }{\pi \hbar }\frac{2b_{n-1}^{2}}{\alpha \sqrt{n-1}
		-\alpha ^{\prime }\sqrt{n}}(\nu _{n,\alpha ^{\prime }}-\nu _{n-1,\alpha })}{
		\frac{w_{1}^{2}\tau ^{2}}{\hbar ^{2}} \left(\frac{\hbar \omega }{w_{1}}-\alpha 
		\sqrt{n-1}+\alpha ^{\prime }\sqrt{n} \right)^{2}+1}  \notag \\
	f_{1}^{\oplus }=\sum_{\substack{ n\geq 0  \\ \alpha \alpha ^{\prime }}}\frac{
		\frac{w_{1}\tau }{\pi \hbar }\frac{2b_{n}^{2}}{\alpha \sqrt{n+1}-\alpha
		^{\prime }\sqrt{n}}(\nu _{n,\alpha ^{\prime }}-\nu _{n+1,\alpha })}{\frac{
		w_{1}^{2}\tau ^{2}}{\hbar ^{2}} \left(\frac{\hbar \omega }{w_{1}}-\alpha \sqrt{n+1}
		+\alpha ^{\prime }\sqrt{n} \right)^{2}+1}  \notag \\
	f_{2}^{\ominus }=\sum_{\substack{ n\geq 2  \\ \alpha \alpha ^{\prime }}}
		\frac{\frac{4c_{n-1}^{2}}{\pi \omega _{c}\tau }\frac{(\nu _{n,\alpha
		^{\prime }}-\nu _{n-1,\alpha })(n-1)}{\alpha \sqrt{n^{2}-n}-\alpha ^{\prime }
		\sqrt{(n-1)(n-2)}}}{\left[\frac{\omega }{\omega _{c}}-\alpha \sqrt{n^{2}-n}
		\! +\! \alpha ^{\prime }\sqrt{(n-1)(n-2)}\right]^{2} + \frac{\tau ^{-2}}{\omega _{c}^{2}}}
		\notag \\
	f_{2}^{\oplus }=\sum_{\substack{ n\geq 1  \\ \alpha \alpha ^{\prime }}}\frac{
		\frac{4c_{n}^{2}}{\pi \omega _{c}\tau }\frac{(\nu _{n,\alpha ^{\prime }}-\nu
		_{n+1,\alpha })n}{\alpha \sqrt{n^{2}+n}-\alpha ^{\prime }\sqrt{n^{2}-n}}}{\left(
		\frac{\omega }{\omega _{c}}-\alpha \sqrt{n^{2}+n}+\alpha ^{\prime }\sqrt{
		n^{2}-n}\right)^{2}+\frac{\tau ^{-2}}{\omega _{c}^{2}}}.  \notag
\end{gather}
Here $w_{1}=\sqrt{2}\hbar \upsilon \lambda _{B}^{-1}$ and $\hbar \omega
_{c}=\hbar ^{2}/m_{2}\lambda _{B}^{2}$ are the characteristic energy scales
for the LL spectra in monolayer ($J=1;\epsilon _{n\alpha }$) and bilayer ($
J=2;\varepsilon _{n\alpha }$) graphene, respectively, and $\alpha ,\alpha
^{\prime }=\pm $ determine whether the corresponding state belongs to the
conduction ($+$) or valence ($-$) band. Also, $\nu _{n,\alpha }$ are the
filling factors of the corresponding LLs, and $b_{0}=1$, $b_{n\geq 1}=1/
\sqrt{2}$ for a monolayer and $c_{0,1}=1$, $c_{n\geq 2}=1/\sqrt{2}$ for the
bilayer.

One can test the selection rules shown in Fig. \ref{fig:transitions}(a) and
the actual polarisation of these transitions
using gated graphene structures \cite{Novoselov,Zhang,QHEbi}. By filling the
monolayer sheet with electrons up to $\nu =2$ (a completely filled $n=0$
LL), one would suppress the intensity of the lowest absorption peak in 
$\boldsymbol{\ell }_{\ominus }$ polarisation and increase the size of the 
$\boldsymbol{\ell }_{\oplus}$ peak. By depleting the monolayer to 
$\nu =-2$ state (emptying the $n=0$ LL) one would achieve the opposite
effect. Similarly, in a bilayer with a completely filled pair of $n=0,1$
LLs which takes place at $\nu =4$, light with $\omega=\sqrt{2}\omega_c$ can be
absorbed only in the $\boldsymbol{\ell }_{\oplus }$ polarisation. By depleting
the bilayer down to $\nu =-4$ one could suppress the absorption in this line
in $\boldsymbol{\ell }_{\oplus }$ polarisation while retaining
$\boldsymbol{\ell }_{\ominus}$ absorption. 

The denser magneto-absorption spectrum in bilayer graphene (as compared to
the monolayer) is one of its features illustrated in Fig.\ref{fig:transitions}
(b). Another feature is that the intensity of the lowest absorption peak
measured in the $\boldsymbol{\ell}_\oplus$ or $\boldsymbol{\ell}_{\ominus}$
polarisations at the filling factor $\nu =0$ may differ for 
different ground states of the bilayer. To illustrate this possibility, let
us compare the absorption spectra for two model ground states of the bilayer.

The first can be attributed to a bilayer with such a large single-particle
splitting between the $n=0$ and $n=1$ states that one of these levels is full
and the other empty, even in the presence of the electron-electron
interaction. This may be due either to the inter-layer assymmetry
\cite{McCann,Others} or caused by the AA/BB intralayer hopping in the last
term of $\hat{H}_2$, Equation \eqref{eq:Ham2}.  Below we refer to such a QHE state as
being antiferromagnetic, stressing that this state is not spin-polarised. 

An alternative form of the ground state can be attributed to the case of a
negligible splitting between the $n=0$ and $n=1$ LLs. In the latter situation
the electron-electron repulsion may lead to the ferromagnetic alignment of
electron spins and the formation a ferromagnetic QHE state with one spin
component of each LL being completely full and the other completely empty.

Since transitions from/to the $n=0$ LL with $\varepsilon _{0}\approx 0$
to/from the states $\varepsilon _{n\pm }$ with $n\geq 2$ are forbidden, 
the intensity and polarisation of the peaks at $\omega=\sqrt{2}\omega_c$
are determined by transitions from/to $n=1$ LL (also with
$\varepsilon_{1}\approx 0$) and directly reflect the occupancy of this state.
If the bilayer ground state is ferromagnetic, with a half-filled $n=1$ LL, the
absorption peak at $\omega=\sqrt{2}\omega _{c}$ will have the same intensity
in both $\boldsymbol{\ell }_{\oplus }$ and $\boldsymbol{\ell }_{\ominus}$
polarisations.  

In contrast, absorption by a $\nu =0$ bilayer with antiferromagnetic ground
state would contain the line at $\omega = \sqrt{2} \omega_c$ only in one
polarisation: in $\boldsymbol{\ell}_{\ominus}$ if $n=1$ LL is empty (fully
occupied $n=0$ LL) and in $\boldsymbol{\ell}_{\oplus }$ if $n=1 $ LL is full.
For comparison, the lowest peak in the spectrum of a monolayer with $\nu =0$
appears in both polarisations, since both transitions to ($\epsilon_{1-}
\rightarrow \epsilon_{0}$) and from ($\epsilon_{0} \rightarrow \epsilon_{1+}$)
half-filled $n=0$ monolayer LL are possible All higher-energy absorption peaks
which involve transitions between filled valence band states ($\alpha =-$) and
empty states in the conduction band ($\alpha =+$) have equal strengths in both
polarisations, which reflect effectively the inter-band nature of these
transitions.  Also, note that a weak transition between $n=0$ and $1$ LLs with
a low frequency $\beta \omega _{c}$ (microwave range) is possible, to the
measure of a small $\beta $ and depending on the LL filling.

We briefly turn our attention to the effect of the electron-electron
interaction on the absorption. Since we have specified that we study the
spectrum for exactly filled levels we can construct magneto-exciton operators
for each case \cite{Halperin}. These operators correspond to collective
excitations of electrons from fully filled to completely empty LLs. We
calculated the shift in the dispersion of such magneoexcitons caused by these
interactions to the first order in the interaction constant
$e^2/(\epsilon_r \lambda_B\hbar\omega_c)$ -- where $\epsilon_r$
is the dielectric constant of graphene -- and find that the Coulomb
interaction has no effect on these excitations when their in-plane wave vector 
$\mathbf{k}$ is zero. This result agrees with recent studies of the
mangetoexciton dispersion in the integer QHE states in monolayer graphene
\cite{Fertig-Brey} and our previous work in the context of semiconductor
heterostructures probed with surface acoustic waves \cite{Smet-SAW}. Therefore
in these cases we can disregard the electron-electron interactions and say
that our conclusions concerning the polarisation properties of FIR absorption
apply to all magnetoexcitons which are symmetric in the spin and valley
indices.

The analysis mentioned above was performed in the lowest order of pertubation
theory in the interaction. Its result does not imply a full Kohn's theorem as
known for electron-electron interactions in a simple parabolic band. In
graphene, the chirality of the carriers is similar to the spin-orbit coupling
in III-V semiconductors in that it plays the role of a non-parabolicity in the
band structure which is known to violate Kohn's theorem in conventional
semiconductors. This means that the conclusions for the lowest order
corrections due to the interaction may not stand for higher orders of
pertubation theory. This issue will be the subject of a seperate investigation
\cite{bib:Ab-falk-unpub}.

Finally, the warping term $\delta \hat{h}_{w}$ in Eq. (\ref{eq:Ham2}) mixes
each state $\chi _{n\alpha ,K}$ with states $\chi _{(n\pm 3)\alpha ,K}$.
This generates weak transitions (with the coefficient $\delta g_2 \sim
(\upsilon_3 m_2 \lambda_B / \hbar)^2 g_2$) between states which are 2 and 4
levels apart. Although such transitions are negligibly weak at high fields
where the first term in $\hat{H}_{2}$ is dominant, they become relevant at
weak fields, where $\lambda _{B}^{-1}\lesssim \gamma _{1}\upsilon
_{3}/\upsilon ^{2}$, and the low-frequency absorption spectrum of a bilayer
aquires an additional structure. In Fig.\ref{fig:transitions}(c) we compare
the absorption spectra in a $\nu =0$ bilayer \cite{broadenning} at $B=1$T
calculated numerically for $\upsilon_{3}=0$ and $\upsilon _{3}=0.2\upsilon$.

In \textit{conclusion}, we have described peculiar optical and FIR
magneto-optical properties of bilayer graphene and compared them with those of
the monolayer material. The zero field optical absorption spectrum shown in
Fig. \ref{fig:AC-zero-field} reflects the existance of four bands in the
electronic spectrum of this material whereas the FIR magneto-optical spectrum
of bilayer graphene, which is much denser than that of a monolayer, reflects
the parabolic dispersion of its low-energy bands $\varepsilon _{c}^{\pm }$.
Due to the interband character of the inter-LL transitions in graphene with
low carrier density, the magneto-optical spectrum in a monolayer and the
higher energy part of the bilayer spectrum ($\omega >3\omega _{c}$) do not
depend on the FIR light polarisation. However, the absorption peak at $\omega
=\sqrt{2}\omega _{c}$ may appear differently in $\boldsymbol{\ell} _{\oplus }$
and $\boldsymbol{\ell} _{\ominus }$ polarised light, depending on the
occupancy of the two zero-energy LL ($n=0$ and $1$) in bilayer graphene
\cite{McCann,QHEbi}. In such spectroscopic studies, the weakness of $g_{2}$
can be overcome once one uses the 2D electrons in graphene in the QHE regime
for the transport detection of FIR absorption.

We thank E. McCann and A. Varlamov for discussions and the EPSRC grant
EP/C511743 for support. This work has been complete during the MPI-PKS
workshop `Dynamics and Relaxation in Complex Quantum and Classical Systems and
Nanostructures', July-October 2006.


\begin{thebibliography}{99}
\bibitem{kpoints} Here, $\mathbf{K}_{\pm }=\pm ({\tfrac{2}{3}} ha^{-1},0)$, $
a$ is the lattice constant.
\bibitem{beta} The sign of $\beta $ is not known, though one can estimate
using bulk graphite parameters \cite{AndoReview} that $|\beta |\lesssim
10^{-2}$.
\bibitem{transv} In contrast to monolayer graphene, a weak absorption of
light polarised perpendicular to the bilayer is possible. A pertubation $
\sigma _{z}eE_{z}d/2$ distinguishes between the on-site energies in the top
and bottom layers separated by spacing $d$, which leads to weak absorption $
g_{2}^{z}=(2\pi e^{2}/\hbar c)f_{2}^{z}$, 
\begin{gather*}
	f_{2}^{z}= a_{z}^{2}\Omega \left[ \tfrac{1}{\Omega +1} + 
		\tfrac{\theta (\Omega -2)}{\Omega -1}\right] ,\quad \Omega \equiv
		\hbar \omega / \gamma _{1}; \\
	f_{2}^{z}(B,\omega )=\frac{a_{z}^{2}}{\pi}\sum_{n\geq 2}\frac{\tau \omega 
		}{\tau ^{2}\omega _{c}^{2}(\frac{\omega }
		{\omega_{c}}-2\sqrt{n^{2}-n})^{2}+1}
\end{gather*}
where the constant $a_{z}=\gamma_{1}d/2\hbar v\sim 10^{-1}$, and the magneto-absorption
spectrum at $\hbar \omega <\frac{1}{4}\gamma _{1}$ involves $\varepsilon
_{n-}\rightarrow \varepsilon _{n+}$ inter-LL transitions.
\bibitem{frequency} For $\hbar \omega \ll \frac{1}{4}\gamma _{1}$ this
result transforms into $f_{2}=1$ suggested by J.Cserti \cite{Czerti} for the
microwave absorption in bilayer graphene. However one should be aware that
Eq. (\ref{g2}) and conclusions of Ref. \cite{Czerti} cannot be applied to $
\hbar \omega \lesssim \epsilon _{\mathrm{L}}=\frac{1}{2}\gamma _{1}(\upsilon
_{3}/\upsilon )^{2}\sim 1$meV. At $\epsilon _{\mathrm{F}}\approx \epsilon _{
\mathrm{L}}$, the warping term $\delta \hat{H}_{w}$ causes the Lifshitz
transition \cite{Lifshitz} in the topology of the Fermi line (in each
valley): from singly-connected at $\epsilon _{\mathrm{F}}>\epsilon _{\mathrm{
L}}$ to four pockets  at $\epsilon _{\mathrm{F}}<\epsilon _{\mathrm{L}}$.
\end{thebibliography}
\end{document}